\documentclass{article}

\usepackage{subfigure}
\usepackage{tabularx}
\usepackage{PRIMEarxiv}
\usepackage{cite}
\usepackage[utf8]{inputenc} 
\usepackage[T1]{fontenc}    
\usepackage{hyperref}       
\usepackage{url}            
\usepackage{booktabs}       
\usepackage{amsfonts}       
\usepackage{nicefrac}       
\usepackage{microtype}      
\usepackage{lipsum}
\usepackage{fancyhdr}       
\usepackage{graphicx}       
\usepackage{amsmath}
\usepackage{amssymb}
\graphicspath{{media/}}     

\title{Graph Contrastive Learning for Multi-Omics Data 
}

\author{
  Nishant Rajadhyaksha,\\
  K.J Somaiya College of Engineering\\
  Mumbai\\
  \texttt{n.rajadhyaksha@somaiya.edu} \\
   \And
  Aarushi Chitkara \\
  St Xaviers College \\
  Mumbai\\
  \texttt{aarushi5812@gmail.com} \\
}

\begin{document}
\maketitle

\begin{abstract}
Advancements in technologies related to working with omics data require novel computation methods to fully leverage information and help develop a better understanding of human diseases. This paper studies the effects of introducing graph contrastive learning to help leverage graph structure and information to produce better representations for downstream classification tasks for multi-omics datasets. We present a learnining framework named Multi-Omics Graph Contrastive Learner(MOGCL) which outperforms several aproaches for integrating multi-omics data for supervised learning tasks. We show that pre-training graph models with a contrastive methodology along with fine-tuning it in a supervised manner is an efficient strategy for multi-omics data classification.
\end{abstract}


\section{Introduction}
Omics, referring to a field of study in biological sciences that ends with -omics, aims at the collective characterization and quantification of pools of biological molecules that translate into the structure, function, and dynamics of an organism or organisms. The use of high throughput biomedical research methods has increased substantially over the past few years such as Whole Genome Sequencing(WGS), RNA sequencing(RNA-seq), chromosome conformation capture (Hi-C) and liquid chromatography coupled with mass spectrometry \cite{Fuentes-Pardo2017-yd,Chen2013-jy}. It is particularly helpful to integrate data from different molecular sources such as microRNA(miRNA),mRNA and DNA methylation to provide insight into the classification and processes of different diseases. Integration of multi-omics data requires efficient computational methods which are capable of correctly modelling and interpreting these relationships. Whilst each omics technology can present only a part of the true biological complexity integrating data from multiple omics technologies can help provide a more universal picture which can help improve results for classification tasks \cite{10.3389/fgene.2019.00166,10.3389/fgene.2017.00084,Canzler2020,Singh2019-tu,https://doi.org/10.1002/sim.6732,metabo12030225,Subramanian2020-ge}.

Several different methodologies have been proposed to help integrate multi-omics data for classification tasks. Generally, different omics data have often been concatenated together or have been subject to ensemble learning where prediction for each omics type is ensembled together \cite{Poirion2021,O'Donnel}. The recent emergence of Graph Neural Networks(GNNs) as an efficient deep learning methodology to model complex systems has prompted researchers to study its effects when paired with multi-omics data \cite{Yin2022-xe,xing2021interpretable}. Graph contrastive learning is a learning paradigm where data-data pairs are utilised to perform discrimination between positive and negative pairs of graph data. Contrastive learning can be used as an effective pre-training strategy for training graphical models on supervised learning tasks. This paper describes a framework named MOGCL which constructs graphs from multi-omics data and utilises contrastive learning as a pre-training strategy to aid downstream classification tasks. We compare our results on benchmark datasets with several different machine-learning methodologies. MOGCL performs better on several metrics such as Area Under the ROC Curve(AUC), Accuracy and F1 score etc.

\section{Literature Review}
\label{sec:headings}

\subsection{Machine learning for multi-omics data}
A significant number of methods have been studied to integrate multi-omics data for various tasks. A large number of methods focus on the semi-supervised integration of multi-omics data without utilising the information from ground truth labels present in the dataset \cite{huang2021integration,9669359,10.1093/bioinformatics/btp543,Tini2019-es}. Most self-supervised methods focus on assigning clusters to different omics groups present in the dataset. With the rapid advancements of biotechnology and the detailed curation of datasets, more labelled samples of phenotypes or traits are being made available for research. This has led to the development of supervised learning models which perform better on multi-omics datasets \cite{Sumo-fil,MENYHART2021949,Tan2020,Costello2018,PICARD20213735}. Kernel methods are powerful machine learning models which operate on a higher dimensional space where linear interpolations can be found between the given data points. Kernel methods are often used as classical machine learning models for analysing and classifying multi-omics. Support Vector Machines(SVM) \cite{Kim2017} and partial least squares \cite{S_Punla2022-vx} are examples of classical machine learning approaches for multi-omics data. Currently, deep learning approaches are commonly adopted for multi-omics integration for various tasks. Islam et al. \cite{Mohaiminul_Islam2020-gf} propose an approach which utilises a convolutional neural network to learn important features for multiple omics types and finally concatenate them to predict breast cancer subtypes. An approach using stacked Sparse Autoencoders (SAE) was proposed by Xu et al. \cite{Xu2019} where representations are produced for each type of omics data using autoencoders which are then fed to a deep flexible neural forest for predicting cancer subtypes.

\subsection{Graph based learning approaches}
Graphs are complex data structures which are used to model many complex phenomena in the real world. Graph Neural Networks (GNN) deal with applying deep learning to graphical data structures. GNNs have several applications such as combinatorial optimizations, neural machine translation, protein-protein interactions, drug discovery \cite{Schuetz2022,peng2021graph,nguyen-etal-2022-multi,mei2021graph,nasiri2021novel,dai2021pike,wang2022advanced,bongini2021molecular}. Recently graph-based approaches have been used for multi-omics integration. Wang et al. \cite{Wang2021} proposed a methodology to convert multi-omics data to graphs and a model named MOGONET consisting of convolutional network (GCN) layers \cite{gcn} to produce refined representations for a downstream classification task on multiple multi-omics datasets whilst also identifying important biomarkers. Xiao et al. \cite{10.3389/fgene.2022.806842} proposed a model named MOGCN for different cancer sub-type classification tasks based on multi-omics data. Graph contrastive learning is an example of a self-supervised training methodology where different graph augmentations are utilised to exploit both structural information and information about features of the dataset to produce better representations. Some common training strategies are pre-training followed by fine-tuning \cite{finetuninggcl} and joint learning \cite{ijcai2021p473}.
Zhu et al. \cite{Zhu:2020vf} proposed a framework titled deep GRAph Contrastive rEpresentation learning (GRACE) which specifically generates two graph views by corruption and attempts to learn node representations by maximizing the agreement of node representations in these two views.

\section{Methodology}
\label{sec:others}
\subsection{Datasets}

We demonstrate the effectiveness of MOGCL by applying our model on two benchmark datasets namely ROSMAP \cite{DeJager2018} which describes information for patients with Alzheimer's Disease (AD) concerning a normal control group (NC) and BRCA which contains data for breast invasive carcinoma (BRCA) PAM50 subtype classification. The omics data used were namely were DNA methylation data (meth), miRNA expression data (miRNA) and mRNA expression data (mRNA). Details about the datasets are further described in table \ref{tab1}.

\begin{table}[h]
\centering
\begin{tabular}{|l|l|l|l|}

\hline
Dataset Name & Labels & \begin{tabular}[c]{@{}l@{}}Number of features for \\ mRNA, meth, miRNA\end{tabular} & \begin{tabular}[c]{@{}l@{}}Number of  features for training \\ mRNA, meth, miRNA\end{tabular} \\ \hline
ROSMAP  & NC: 169, AD: 182 & 55,889, 23,788, 309 & 200, 200, 200 \\ \hline
BRCA    & \begin{tabular}[c]{@{}l@{}}Normal-like: 115, \\ Basal-like: 131, \\ HER2-enriched: 46, \\ Luminal A: 436, \\ Luminal B: 147\end{tabular} & 20,531, 20,106, 503                                                                 & 1000, 1000, 503                                                                               \\ \hline
\end{tabular}
\caption{Summary of datasets}

\label{tab1}
\end{table}

\subsection{Contrusting graphs from multi-omics data}

In this section, we describe the methodology of converting multi-omics data to a graphical structure which can then be leveraged by powerful graph neural network models for further processing. Our task can be defined as defining graphs \begin{math} G \end{math} = \begin{math} (V, E) \end{math} where \textit{V}, \textit{E} represent vertices and edges of the graph respectively. We utilise the feature matrices we obtain after preprocessing each type of omics data. The feature matrix for each omics type is represented as  \begin{math} X \in \mathbb{R}^{n \times d } \end{math} where for the ROSMAP dataset d is 200 for each of the omics types and n is 351. Similarly for the BRCA dataset n is 875 and d ranges from 1000 for mRNA and meth data to 503 for miRNA data. The nodes V of graph G represent the users from which the omics data is collected. We construct an adjacency matrix \begin{math} A \in \mathbb{R}^{n \times n } \end{math} to represent G with each element in the adjacency matrix representing a node. We denote a weighted edge from node i to node j of the graph as the element present at the i\textsuperscript{th} row and j\textsuperscript{th} column of A. Such an adjacency matrix is constructed for each type of omics data respectively. A pairwise distance matrix is constructed for data for the points of the particular omics dataset using cosine similarity \cite{7577578} as the distance metric. The distance between node i and node j is denoted by $t_{ij}$. A parameter k is introduced which represents the number of edges to be formed per node. An adjacency parameter is then chosen by selecting the \begin{math} n\times{k}^{th} \end{math} value from a sorted array of pairwise distances between all data points. Edges E is then selected on the criteria of the distance between data points being smaller than the adjacency parameter. This ensures that the number of edges per node is k. A weight of \begin{math} 1 - t_{ij} \end{math} is assigned to the edge from node i to node j if belongs to the set of selected edges. An adjacency matrix is prepared for each of the omics types present in the respective dataset by following the methodology described above.

\subsection{Graph constrative learning}

In this section, we describe our training methodology which utilises graph contrastive learning. We use GRACE \cite{Zhu:2020vf} which serves as our self-supervision model. Our contrastive learning methodology consists of two stages namely i) data augmentation and ii) contrastive learning. Augmentation functions such as removing random edges and feature masking are used to create augmented views of a graph. For augmenting edges, we randomly remove a portion of edges from the original graph. We sample a random masking matrix $\Tilde{R} \in \{0,1\}^{N\times{N}}$. Where the elements of R are drawn from a Bernoulli distribution $\Tilde{R} \sim Bern(1-pr)$ where pr is the probability of each edge is removed. We choose pr to be 0.4 for our study. The resulting adjacency matrix can be given as \begin{math}\Tilde{A} = A \circ{\Tilde{R}}\end{math}. For augmenting features we randomly mask the dimensions of a feature vector by replacing them with zeros. We sample random feature vectors to construct a matrix \begin{math}\Tilde{M} \in \{0,1\}\end{math} according to a Bernoulli distribution having a similar size as feature matrix \emph{X}. The augmented feature matrix can then be represented by \begin{math}\Tilde{X} = X \circ{\Tilde{M}}\end{math}. We use a GCN \cite{gcn} model as an encoder model which helps represent the augmented views of a given graph and denote it with \emph{f}. Let \begin{math} U = f(\Tilde{X_{1}},\Tilde{A_{1}}) \end{math} and \begin{math} V = f(\Tilde{X_{2}},\Tilde{A_{2}}) \end{math} be the representations generated after processing two graphs with our shared encoder model. We aim to maximise the agreement between similar nodes in the latent space and minimise the agreement between the rest of the contrasting nodes. To achieve this we make use of the Normalized Temperature-scaled Cross Entropy Loss (NT-Xent) \cite{simclr}. NT-Xent loss is given by eq \ref{eqn1}.

\begin{equation}
\label{eqn1}
    \ell (\boldsymbol{u}_i, \boldsymbol{v}_i) =
\log \frac {e^{\theta\left(\boldsymbol{u}_i, \boldsymbol{v}_{i} \right) / \tau}} {e^{\theta\left(\boldsymbol{u}_i, \boldsymbol{v}_{i} \right) / \tau} + \displaystyle\sum_{k \neq i} e^{\theta\left(\boldsymbol{u}_i, \boldsymbol{v}_{k} \right) / \tau} + \displaystyle\sum_{k \neq i}e^{\theta\left(\boldsymbol{u}_i, \boldsymbol{u}_k \right) / \tau}},
\end{equation}

where $u_{i}$ and $v_{i}$ represent the $i^{th}$ feature vector from the feature matrix \emph{U} and \emph{V} respectively. $\tau$ represents a temperature parameter. $\theta$ is a similarity function given in equation \ref{eq2}.

\begin{equation}
\label{eq2}
    \theta(u,v) = c(n(u),n(v))
\end{equation}

where c(.,.) is the cosine similarity function and n(.) represents any non-linear function such as ReLU \cite{relu} or LeakyReLU \cite{LeakyRelU} etc. We finally optimise the weights of the shared encoder model on the NT-Xent loss.

\begin{figure}[htp]
    \centering
    \includegraphics[scale=0.8]{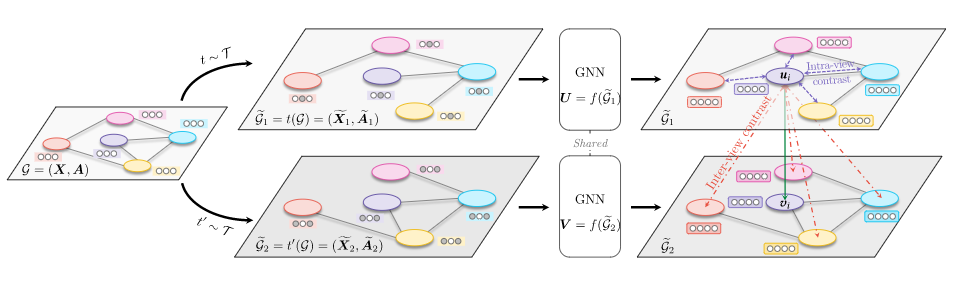}
    \caption{Contrastive Learning for GNN Encoder}
    \label{fig:Contrastive Learning for GNN ENcoder}
\end{figure}

The GCN encoder is further trained in a supervised manner using labels from the given dataset. The encoder model was trained for a downstream classification task using pre-training followed by fine-tuning. In pre-training, we first fully train an encoder model for each omics type in an unsupervised manner. We later fine-tune the models using label information from the given dataset. Let \begin{math} \Tilde{f} \end{math} be the pre-trained GCN encoder. We utilise linear layers in conjunction with concatenated features produced from the encoder models to produce predicted label \begin{math} \Tilde{Y} = \Tilde{f}(X,A) \end{math}. We use the Cross-Entropy Loss to calculate the loss for predicted labels $\Tilde{Y}$ and true labels $Y$ and finally optimise our encoder model on this loss.
\begin{figure}[htp]
    \centering
    \includegraphics[scale=0.8]{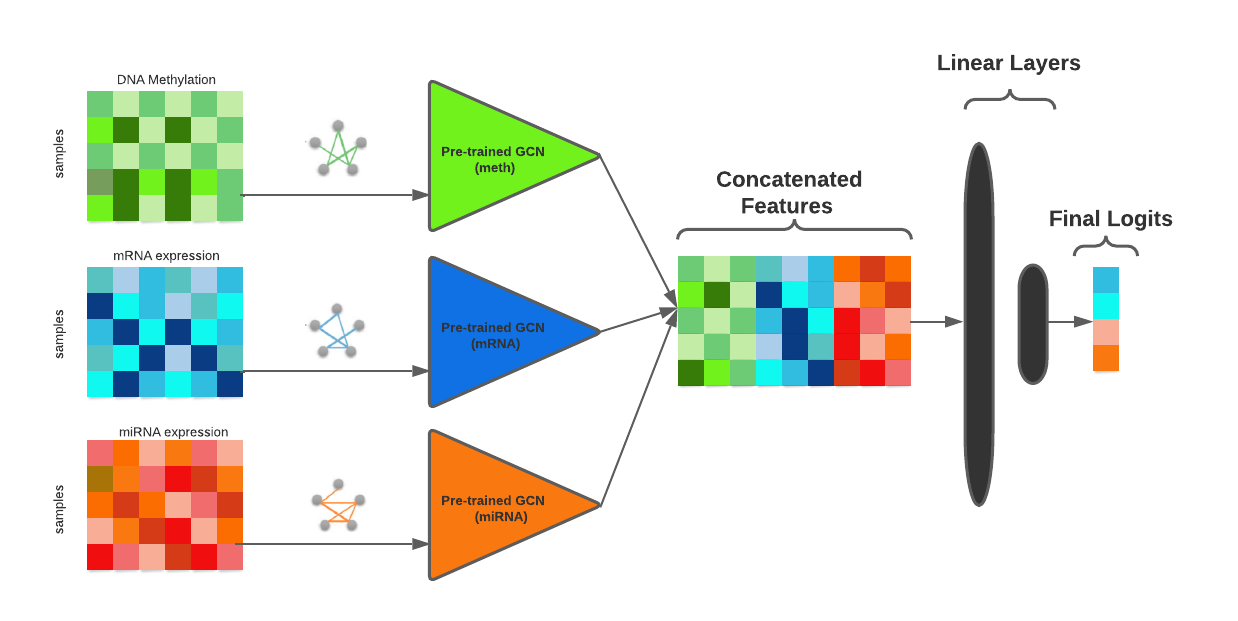}
    \caption{Downstream Supervised Training of GNN Encoder}
    \label{fig:GNN ENcoder}
\end{figure}

\subsection{Experiments}

In this section we describe the experiments we perform to evaluate our MOGCL framework. We first produce graphs for each omics type in our datasets and train a separate encoder model for each one respectively. We finally concatenate the features produced by each encoder model and train the encoder model in a pre-training followed by fine-tuning methodology. We compare our classification results to the ones described in \cite{Wang2021} to evaluate the efficiency of introducing a contrastive learning methodology for the given classification task. Performance of all permutations of encoder models is calculated by conducting \begin{math} r=5 \end{math} runs with random weight initialisations for each permutation. We measure the performance of our model on metrics such as accuracy, f1-score and AUC for the ROSMAP dataset and use accuracy,f1-weighted score and f1-macro score to evaluate the BRCA dataset. We use PyTorch-geometric \cite{Fey/Lenssen/2019}, PyGCL \cite{Zhu:2021tu} and pytorch-lightning\cite{falcon2019pytorch} for conducting our experiments. Adam \cite{adam} optimizer with a learning rate of 0.0001 is utilised for all our experiments.  We use a two-layered GCN as our encoder model which is used in graph contrastive learning. We further use two linear layers in conjunction with our encoder model to perform fine-tuning with the given true labels. We compress all feature vectors to a 100-dimensional latent space for all our experiments. We try to visualise the effects of our pretraining strategy by visualising the feature vectors before and after processing them with our encoder models for each omics type. t-SNE \cite{vanDerMaaten2008} was utilised to compress feature vectors to a two-dimensional space in order to produce visualisations.

\section{Results and Discussion}

The results for the classification task for ROSMAP and BRCA datasets are displayed in table \ref{results_ROSMAP} and table \ref{RESULTS_BRCA} respectively. 

\begin{table}[h]

\caption{Results for classification task on ROSMAP}
\centering
\begin{tabular}{|l|l|l|l|}
\hline
Method       & Accuracy       & F1           & AUC           \\ \hline
KNN          & 0.657 ±  0.036 & 0.671 ± 0.044 & 0.709 ± 0.045 \\ \hline
SVM          & 0.770 ± 0.024  & 0.778 ± 0.016 & 0.770 ± 0.026 \\ \hline
Lasso        & 0.694 ± 0.037  & 0.730 ± 0.033 & 0.770 ± 0.035 \\ \hline
RF           & 0.726 ± 0.029  & 0.734 ± 0.021 & 0.811 ± 0.019 \\ \hline
XGBoost      & 0.760 ± 0.046  & 0.772 ± 0.045 & 0.837 ± 0.030 \\ \hline
NN           & 0.755 ± 0.021  & 0.764 ± 0.021 & 0.827 ± 0.025 \\ \hline
GRridge      & 0.760 ± 0.034  & 0.769 ± 0.029 & 0.841 ± 0.023 \\ \hline
block PLSDA  & 0.742 ± 0.024  & 0.755 ± 0.023 & 0.830 ± 0.025 \\ \hline
block sPLSDA & 0.753 ± 0.033  & 0.764 ± 0.035 & 0.838 ± 0.021 \\ \hline
NN\_NN       & 0.766 ± 0.023  & 0.777 ± 0.019 & 0.819 ± 0.017 \\ \hline
NN\_VCDN     & 0.775 ± 0.026  & 0.790 ± 0.018 & 0.843 ± 0.021 \\ \hline
MOGONET\_NN  & 0.804 ± 0.016  & 0.808 ± 0.010 & 0.858 ± 0.024 \\ \hline
\textbf{MOGCL (ours)} & \textbf{0.818 ± 0.014}  & \textbf{0.818 ± 0.014}       & \textbf{0.866 ± 0.021}       \\ \hline
\end{tabular}

\label{results_ROSMAP}
\end{table}

\begin{table}[h]

\caption{Results of classification task on BRCA.}
\centering
\begin{tabular}{|l|l|l|l|}
\hline
Method       & Accuracy       & F1-Weighted           & F1-Macro           \\ \hline
KNN          & 0.742 ±  0.024 & 0.730 ± 0.023 & 0.682 ± 0.025 \\ \hline
SVM          & 0.729 ± 0.018  & 0.702 ± 0.015 & 0.640 ± 0.017 \\ \hline
Lasso        & 0.732 ± 0.012  & 0.698 ± 0.015 & 0.642 ± 0.026 \\ \hline
RF           & 0.754 ± 0.009  & 0.733 ± 0.010 & 0.649 ± 0.013 \\ \hline
XGBoost      & 0.781 ± 0.008  & 0.764 ± 0.010 & 0.701 ± 0.017 \\ \hline
NN           & 0.754 ± 0.028  & 0.740 ± 0.034 & 0.668 ± 0.047 \\ \hline
GRridge      & 0.745 ± 0.016  & 0.726 ± 0.019 & 0.656 ± 0.025 \\ \hline
block PLSDA  & 0.642 ± 0.009  & 0.534 ± 0.014 & 0.369 ± 0.017 \\ \hline
block sPLSDA & 0.639 ± 0.008  & 0.522 ± 0.016 & 0.351 ± 0.022 \\ \hline
NN\_NN       & 0.796 ± 0.012  & 0.784 ± 0.014 & 0.723 ± 0.018 \\ \hline
NN\_VCDN     & 0.792 ± 0.010  & 0.781 ± 0.006 & 0.721 ± 0.018 \\ \hline
MOGONET\_NN  & 0.805 ± 0.017  & 0.782 ± 0.030 & 0.737 ± 0.038 \\ \hline
\textbf{MOGCL (ours)} & \textbf{0.853 ± 0.005}  & \textbf{0.851 ± 0.010}       & \textbf{0.823 ± 0.006}       \\ \hline
\end{tabular}
\label{RESULTS_BRCA}
\end{table}
The performance of MOGCL is compared with the following classification algorithms 1) K-nearest neighbour classifier (KNN). K-nearest neighbours are chosen from the training data to make label predictions during evaluation. 2) Support Vector Machine classifier (SVM). 3) Lasso which is L1-regularised linear regression. A unique model was trained to forecast each class's probability in Lasso, and the class with the greatest foretasted probability was chosen as the final prediction of the model's overall class label 4) Random Forest classifier (RF). 5) Extreme Gradient Boosting (XGBoost) is a distributed, scalable gradient-boosted decision tree (GBDT) machine learning framework. 6) Fully connected Neural Network (NN) classifier. loss for the fully connected NN was calculated by the cross-entropy loss. 7) Adaptive group-regularized ridge regression (GRridge). 8) block PLSDA mentioned in DIABLO \cite{Singh2019-tu}. block PLSDA performs latent Discriminant Analysis (LDA) to project multi-omics data to a latent space. To categorise a discrete outcome, block PLSDA integrates various omics data types measured on the same set of samples. 9) block sPLSDA. 10) MOGONET\_NN. MOGONET\_NN is architecturally similar to MOGCL but does not use a pre-training strategy. 
We achieve significant results by following our pre-training methodology as it performs better than the other models on all metrics used to measure the results. For the ROSMAP dataset MOGCL achieves an average accuracy of 0.818 in comparison to 0.804 achieved by MOGONET\_NN. following this trend MOGCL achieves an F1-score and AUC of 0.818 and 0.866 as compared to 0.808 and 0.856 achieved by MOGONET\_NN. For the BRCA dataset MOGCL achieves an accuracy of 0.853 as compared to 0.805 for MOGONET\_NN. MOGCL receives an F1-weighted score of 0.851 and an F1-macro score of 0.823 as compared to 0.782 and 0.737 respectively for MOGONET\_NN. This demonstrates that adopting a graph based semi-supervised learning strategy in addition to fine-tuning for a downstream task is an effective training strategy for training models on multi-omics datasets. 

\begin{figure}[htp]
    \centering
    \includegraphics[scale=0.6]{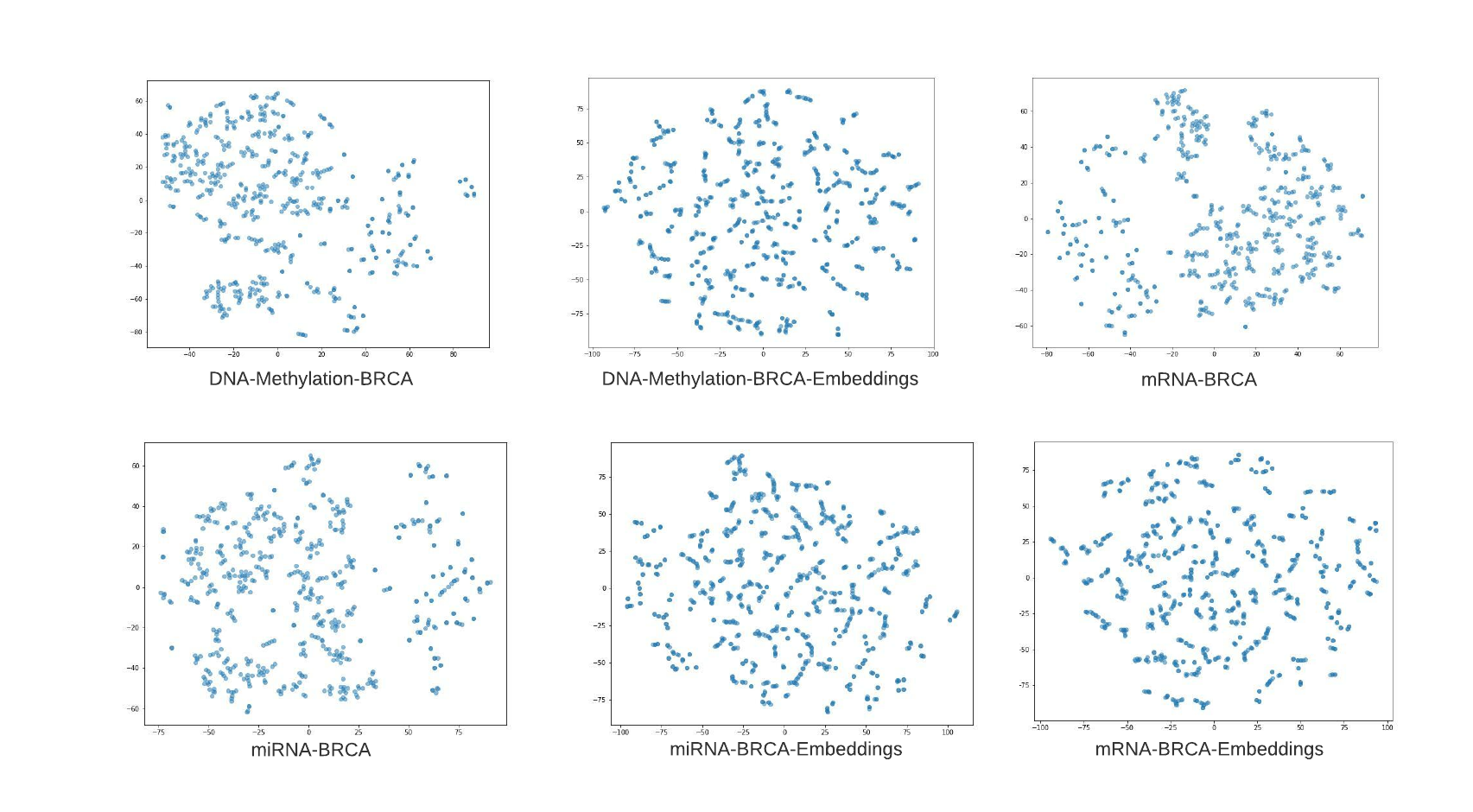}
    \caption{BRCA Embeddings}
    \label{fig:BRCA Embeds}
\end{figure}

\begin{figure}[htp]
    \centering
    \includegraphics[scale=0.65]{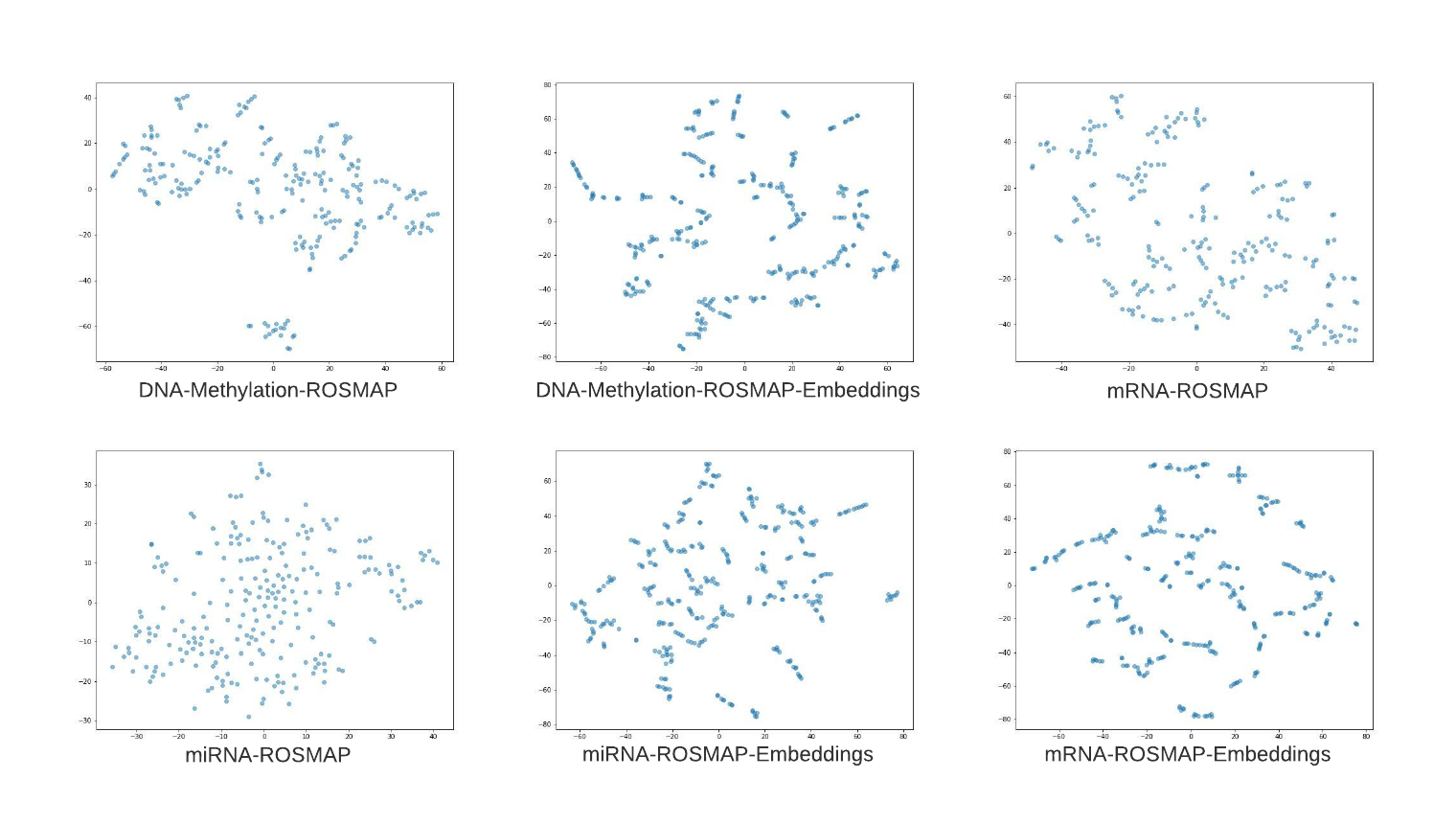}
    \caption{ROSMAP Embeddings}
    \label{fig:ROSMAP Embeds}
\end{figure}

We demonstrate the effects of adopting a semi-supervised methodology of training by analysing \ref{fig:BRCA Embeds} and \ref{fig:ROSMAP Embeds}. We visualise the feature matrices \emph{X} by projecting data points into a two-dimensional plane by utilising t-SNE. Similarly, we map the feature vectors produced by the GCN encoders to a 2-dimensional space and compare the results. MOGCL tries to cluster embeddings in the absence of labels to create more structured representations during the pre-training phase. Better representation help during the fine-tuning phase which in turn helps produce better classification scores.

\begin{figure}[htp]
    \centering
    \includegraphics[scale=0.6]{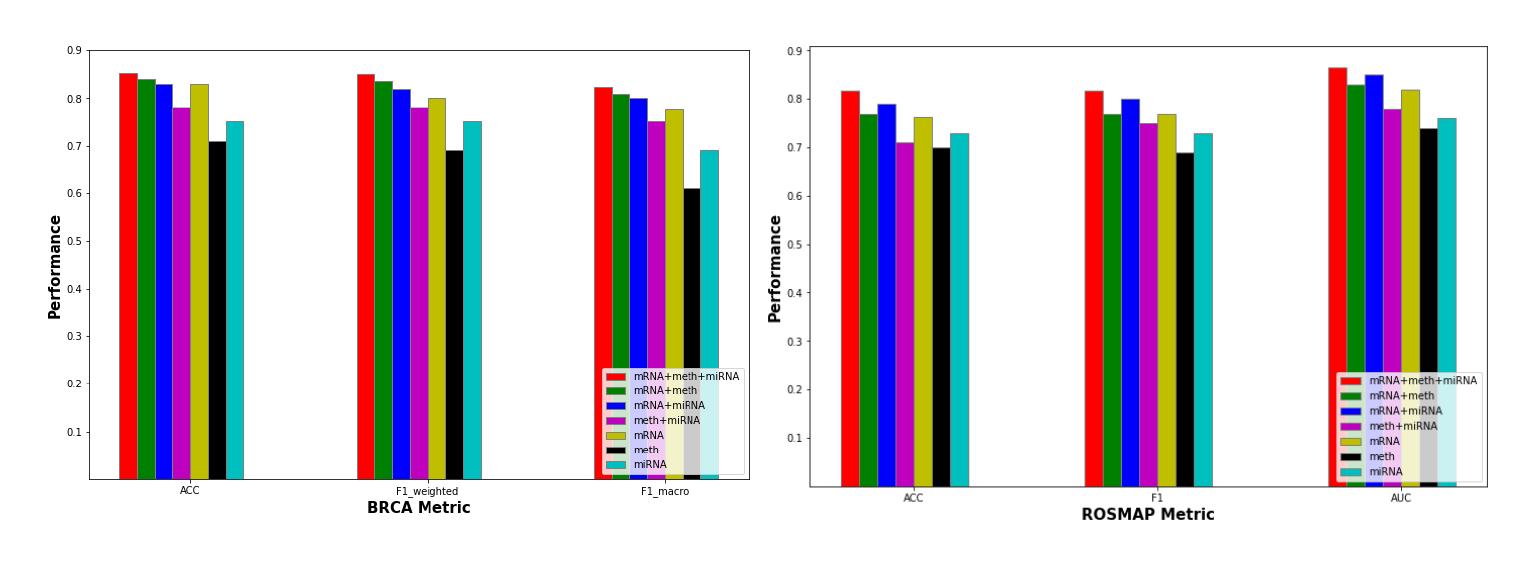}
    \caption{Performance of Omics-Types}
    \label{fig:BRCA bars}
\end{figure}

Figure \ref{fig:BRCA bars} represents the performance of permutation of different omics types when processed by MOGCL. We pre-train three encoder models for all omics types in the study respectively. To calculate performance we select a permutation of these encoder models and train them using true labels in a supervised manner. MOGCL performs its best when fed information by concatenating all the omics types together for both the ROSMAP and BRCA datasets. For the BRCA dataset, a combination of mRNA and DNA-Methylation data provides the next best results however for the ROSMAP dataset a combination of mRNA and miRNA provides the next best set of results. For both the ROSMAP and BRCA datasets using only a single omics type provides the worst results. Using only DNA-Methylation data is the least useful option followed by miRNA and mRNA data across both BRCA and ROSMAP datasets.

\section{Conclusion}
This paper introduces a novel framework named MOGCL which introduces a graph contrastive learning methodology for multi-omics data classification. We first provide a comprehensive literature survey regarding work done in the field of machine learning relating to graph-based learning methods and multi-omics data. A method for constructing graphs from multi-omics data is discussed. We then describe our framework MOGCL which uses GRACE as a pre-training method followed by fine-tuning with true labels in a supervised setting. We discuss our results for the BRCA and ROSMAP datasets and show that our framework performs better than other baselines used for this study. The use of permutations of different omics types is discussed by analysing performance across different metrics. We discuss the effects of adopting a semi-supervised pre-training strategy by visualising the embeddings produced by our graph encoders. We finally conclude that adopting a pre-training methodology is an efficient way to train graphical models for classification problems involving multi-omics datasets. Future works could include experimenting with different contrastive learning methodologies to determine which one is the most efficient. Experiments can be conducted for different GNNs such as Graph Attention Networks (GAT) or Graph Isomorphism Networks (GIN) etc. to determine which one can serve as the best encoder for supervised learning on multi-omics datasets.

\bibliographystyle{unsrt}  
\bibliography{references}

\end{document}